\input harvmac
\def \OO {{\cal O}}
\def \SYM {{\rm SYM}}
\def \De {{\Delta}}

\def \tt {\tau}

\def \six {{\textstyle {1 \ov 6}}}
\def \A {{\cal A}}

\def\ra{\rightarrow}
\def \ads{AdS$_5$\ }

\def \four{{\textstyle {1\ov 4}}}
 \def \third { \textstyle {1\ov 3}}

\def \ep{\epsilon}
\def\D{{\cal D}}

\def \N {{\cal N}}
\def \C {{\cal C}}

\def \k {\kappa} 

\def \g {\gamma}
\def \del {\partial}

\def \na {\nabla}

\def \ha{{\textstyle{1\over 2}}}
\def \na {\nabla }
\def \D {\Delta}
\def \a {\alpha}
\def \b {\beta}
\def\r {\rho}
\def \s {\sigma}
\def \p {\phi}
\def \m {\mu}
\def \n {\nu}
\def \vp {\varphi }

\def \Ga {{\Gamma}}
\def \td {\tilde }
\def \d {\delta}

\def \C {{\cal C}}

\def \inv {^{-1}}
\def \ov {\over }
\def \four{{\textstyle{1\over 4}}}

\def \eight{{\textstyle{1 \ov 8}}}
\def \D {{\cal D}} 
 \def \KK {{\cal K}} 
 \def \dM {{\del M}}
 \def \vex {{\vec{x}}}

\def \lam {{\lambda}}
\def \hh {{\hat{h}}}

\def \dM {{\del M}}
\def \cL {{\cal L}}

\def \hx {{\hat{x}}} 

\def \he {{\rm h}}

 \def \d {\delta} 
\def \L {{\Lambda}}

\def \ge {{\cal G}}
\def \A {{\cal A}}
\def \l {\lambda}
\def \YM {{\rm YM}}
\def \lr { \lref}
\def\np {{  Nucl. Phys. }}
\def \pl {{  Phys. Lett. }}

\def \pr  {{ Phys. Rev. }}

\def \KK {{\cal K}}

\baselineskip8pt
\Title{\vbox
{\baselineskip 6pt
{\hbox {Imperial/TP/97-98/039}}{\hbox{ NSF-ITP-98-049 }}
{\hbox{hep-th/9804083}} 
{\hbox{   }}
}}
{\vbox{
\centerline {$D=4$ Super Yang-Mills, 
$D=5$ gauged supergravity, }
\centerline { } 
 \centerline { and $D=4$ conformal supergravity }
\vskip4pt }}

\centerline  { Hong Liu$^1$\footnote {$^*$} 
{e-mail address: hong.liu@ic.ac.uk} 
and 
A.A. Tseytlin$^{1,2}$\footnote{$^{\star}$}
{\baselineskip8pt e-mail address:
tseytlin@ic.ac.uk}\footnote{$^{\dagger}$}{\baselineskip8pt
Also at  Lebedev  Physics
Institute, Moscow.} 
}



\medskip

\medskip
\smallskip\smallskip
\centerline {$^1$\it  Theoretical Physics Group, Blackett Laboratory,}
\smallskip
\centerline {\it  Imperial College,  London SW7 2BZ, U.K.}
\medskip
\centerline {$^2$\it Institute of   Theoretical Physics, University of
California,}
\smallskip
\centerline {\it  Santa Barbara, CA93106, USA}
\bigskip\bigskip
\centerline {\bf Abstract}
\medskip
\baselineskip10pt
\noindent
\medskip
We consider the role of $\N=4$ conformal supergravity in the 
duality relation between  $\N=4$  SYM theory and $D=5$ gauged
supergravity expanded near the Anti de Sitter background. 
We discuss the structure of the SYM effective action in the conformal 
supergravity background, in particular,  terms related to conformal anomaly.
Solving the leading-order Dirichlet problem for the metric
perturbation in AdS background we explicitly compute the
bilinear graviton term in the $D=5$ Einstein action, demonstrating
its equivalence to the linearized Weyl tensor squared part
of the gravitational effective action induced by SYM theory. 
We also compute the  graviton-dilaton-dilaton 
3-point function which is found to have the form consistent 
with conformal invariance of the boundary theory.

\Date {April 1998}

\noblackbox \baselineskip 15pt plus 2pt minus 2pt 
\lr\deal{S. P. de Alwis, ``Supergravity, DBI action and black hole physics",
hep-th/9804019.}

\lr\dps{M.R. Douglas, J. Polchinski and A. Strominger,
 ``Probing 5-dimensional black holes with
 D-branes", JHEP 12 (1997) 003, 
 hep-th/9703031.}
\lr \hog {G.T. Horowitz and H. Ooguri, ``Spectrum of large n gauge
theory from supergravity",
hep-th/9802116.}
\lr \gkp {S.S. Gubser, I.R.  Klebanov and A.M.  Polyakov,
``Gauge theory correlators from non-critical string theory",
hep-th/9802109.}
\lr \wit {E.  Witten, ``Anti de Sitter space and holography", 
hep-th/9802150.}
\lr \ffz{ S. Ferrara,   C. Fronsdal and A. Zaffaroni, 
``On N=8 Supergravity on $AdS_5$ and $N=4$
Superconformal Yang-Mills theory", 
hep-th/9802203}
\lr\rig{ R.J. Riegert, ``A non-local action for the trace anomaly",
\pl B134 (1984) 56.}
\lr\polya{ A.M. Polyakov, ``Quantum geometry of bosonic strings",
 \pl B103 (1981) 207.}

\lr \ktv{ M. Kaku,  P.K.  Townsend  
and P. van Nieuwenhuizen, ``Properties of
conformal supergravity",
\pr D17 (1978) 3179. }
\lr \kl { I.R. Klebanov, ``World-volume approach
to absorption by non-dilatonic branes",
\np B496 (1997) 231, hep-th/9702076.}
\lr \gkt{S.S. Gubser, I.R.  Klebanov and A.A.
  Tseytlin, ``String theory and classical
  absorption by three-branes", 
  \np B499 (1997) 41, hep-th/9703040. }
\lr \gk{S.S. Gubser and I.R. Klebanov,
 ``Absorption by branes and Schwinger
terms in the world-volume
theory", Phys. Lett. B413 (1997) 41, 
hep-th/9708005.}

\lr\ferandr{
L. Andrianopoli and S. Ferrara, ``K-K excitations
on $AdS_5 \times S^5$
as $N=4$ ``primary'' superfields", {
hep-th/9803171}.}

\lr \pol{ A.M. Polyakov, ``String theory and
quark confinement",
hep-th/9711002.}
\lr\poll{ A.M. Polyakov, ``A few projects in
string theory", Les Houches Sum. Sch. 1992:783, 
hep-th/9304146. }

\lr\bdd{ E. Bergshoeff, M.  de Roo  and B. de Wit, 
``Extended conformal supergravity", 
\np B182 (1981) 173. }
\lr \mald { J. Maldacena, ``The large $N$ limit of
superconformal
field theories and supergravity", {
hep-th/9711200}.}

\lr\chep{ I. Chepelev and A.A.  Tseytlin, 
``Long-distance interactions of branes: correspondence between
       supergravity
       and super Yang-Mills descriptions", hep-th/9709087.
       }

\lr \malda {J.M. Maldacena,  ``Branes probing black holes",
hep-th/9709099.} 

\lr \ft {E.S. Fradkin and A.A. Tseytlin, ``Conformal Supergravity", 
Phys.  Rep. 119 (1985) 233. }
\lr\fftt{ E.S. Fradkin  and A.A. 
 Tseytlin, ``Conformal anomaly in Weyl theory and
 anomaly free superconformal theories", 
 \pl B134 (1984) 187. }
\lr \ftt{E.S. Fradkin  and A.A.  Tseytlin,
``One-loop beta function in conformal
supergravity", 
 \np B203 (1982) 157.   }
\lr \ste{ P.S. Howe, K.S. Stelle and
  P.K. Townsend, ``Miraculous ultraviolet
  cancellations in supersymmetry made manifest", 
\np B236 (1984) 125.}
\lr\ttss{
A.A. Tseytlin, 
``Mobius infinity subtraction and 
effective action  in sigma model approach to  closed string  theory",
\pl B208 (1988) 221.}

\lr\bss{ E. Bergshoeff,  A. Salam   and  E. Sezgin, 
``Supersymmetric $R^2$ actions, conformal invariance and Lorentz Chern-Simons
term in 6 and 10 dimensions",
\np B279 (1987) 659.}

\lr\raj{A. Rajaraman, ``Two-form fields and 
the gauge theory description of
black holes", 
hep-th/9803082.}
     
\lr\hashi{
S.S. Gubser, A. Hashimoto, I.R. Klebanov  and  M.
       Krasnitz,
       ``Scalar absorption and the breaking of the world volume conformal
       invariance",
       hep-th/9803023.}

\lr \dero{M. de Roo, ``Matter coupling in N=4
supergravity", 
 \np B255 (1985) 515; 
 M. de Roo and  P. Wagemans, ``Gauge matter coupling in N=4
supergravity", 
 \np B262 (1985) 644.}
 
\lr \rv {H. R\"omer and P. van Nieuwenhuizen,
  ``Axial anomalies in $N=4$ conformal supergravity",
  \pl B162 (1985) 290.}

\lr \duff{M.J. Duff, ``Observations on conformal anomalies",
\np B125 (1977) 334; ``Twenty years of the Weyl anomaly", 
Class. Quant. Grav. 11 (1994) 1387, hep-th/9308075.} 
\lr \fv {E.S. Fradkin  and G.A. Vilkovisky, ``Conformal off mass 
shell  extension and elimination of conformal anomalies", 
\pl B73 (1978)  209.} 

\lr\vil{ A.O. Barvinsky, Yu.V. Gusev, G.A. Vilkovisky and V.V. Zhytnikov, 
``The one-loop effective action and trace anomaly in four dimensions", 
Nucl. Phys. B439 (1995) 561, 
hep-th/9510037;
A.O. Barvinsky, A.G. Mirzabekian   and  V.V. Zhytnikov, 
  ``Conformal decomposition of the effective action and covariant
       curvature
       expansion",  gr-qc/9510037.  }
       
   \lr \frap{E.S.  Fradkin and M.Ya. Palchik,
   ``Conformal Quantum Field Theory in D
   dimensions"  
   (Kluwer, Dordrecht, 1996).}
   
   \lr \arv{  I.Ya. Aref'eva and  I.V. Volovich,
   ``On large N conformal theories, field theories in Anti de Sitter space  and
   singletons",
   hep-th/9803028. }
   
\lr\hawkgib{G.W. Gibbons and S.W. Hawking, ``Action integrals 
and partition functions in quantum gravity", \pr D15 (1977) 2752.}

   \lr\gaug { M. G\"unaydin, L.J. Romans and N.P.
   Warner, 
   ``Gauged N=8 supergravity in five 
   dimensions", 
   \pl B154 (1985) 268;
   M. Pernici, K. Pilch and  P.
   van Nieuwenhuizen, ``Gauged N=8 D=5
    supergravity",  Nucl. Phys. B259 (1985)
    460.
    }
   \lr\bed{E. Bergshoeff, M.  de Roo  and B. de Wit, 
``Conformal supergravity in ten    
dimensions",  \np B217 (1983) 489. }
   
   \lr\osb{J. Erdmenger and H. Osborn, ``Conserved currents and the
   energy-momentum tensor in conformally invariant theories for general
   dimensions", \np B483 (1997) 431, 
   hep-th/9605009.}
   
\lr\freed{D.Z. Freedman, S.D. Mathur, A. Matusis and L. Rastelli, 
``Correlation functions in the CFT$_d$/AdS$_{d+1}$
correspondence", hep-th/9804058.}
\lr\muck{
W. M\" uck and K.S. Viswanathan, 
``Conformal field theory correlators from classical scalar field theory on AdS$_{d+1}$", 
hep-th/9804035.}

\lr\nensfet{
M. Henningson and  K. Sfetsos,
``Spinors and the AdS/CFT
correspondence",  hep-th/9803251. }

\lr\hawkgib{G.W. Gibbons and S.W. Hawking, ``Action integrals 
and partition functions in quantum gravity", \pr D15 (1977) 2752.}

\lr \kallo{
P. Claus, R. Kallosh, J. Kumar, P. Townsend and A.Van Proeyen,
``Conformal theory of M2, D3, M5 and D1+D5 branes'', {
hep-th/9801206}.}

\newsec{Introduction}
The aim of  this  paper  is to try 
to  check and    clarify  further the recent 
proposal \refs{\gkp,\wit}  about the relation    between the 
 generating functional  for  correlators 
 of marginal operators 
 of large $N$ four-dimensional $\N=4$ super Yang-Mills (SYM) 
  theory and 
 the classical action  of  five-dimensional 
 $\N=8$ gauged supergravity (GSG) expanded near AdS$_5$ vacuum 
and evaluated on  the  solutions 
 with  the Dirichlet  boundary  conditions.  This is 
a  specific realisation of the  duality conjecture  of \mald\ 
(based on  earlier work 
of \refs{\kl,\gkt,\gk,\pol,\dps}).

  One of  the  important  points of our 
  discussion  will be the  role  of  
$D=4, \N=4$  conformal supergravity
 (CSG) \refs{\ktv,\bdd,\ft}  in the relation between  the 
$D=4, \N=4$ SYM and $D=5,\N=8$ GSG theories.\foot{Though the   conformal
supergravity 
 does not  appear 
 directly from string theory and so far  
is  only  a  formal `bridge' 
between SYM and GSG theories, let us still 
 recall that  the 
$\N=4$ CSG (coupled to four  $\N=4$ SYM multiplets) 
is  remarkable  in   being 
 a   unique 
locally  superconformal 4-dimensional  theory 
which is ultraviolet-finite \refs{\ftt,\ft,\ste} 
 and thus 
(conformal and axial) anomaly free \refs{\fftt,\rv}.}
 The 
 relevance
of  the  conformal supergravity
in this context was  already  noted in  \ffz.


Coupling  SYM theory  to CSG multiplet and integrating 
over the SYM  fields in a way preserving general covariance
one finds  the  effective action $W$ which depends 
on the CSG fields as well as on the fields of the 
conformal anomaly multiplet.
If one is interested only in relating the
 derivatives of  $W$ 
to the correlation functions of  marginal operators of
SYM theory viewed as a conformal theory in flat space,  
the terms involving  anomalous  degrees of freedom 
 may  be separated out and ignored. 
We  believe, however, that there is a broader 
picture    which goes beyond the  correspondence between 
correlators of the boundary conformal theory and 
GSG action in the  AdS background 
in which the partition function $Z=e^{-W}$  of the SYM
theory in a supergravity background
should be  given a more fundamental interpretation than just 
a  formal sum  of conformal field theory correlators
multiplied by auxiliary sources. 
In particular (by analogy with  familiar 2d case)
 $Z$ should be computed in a way preserving 
general (super)covariance and thus including
non-linear couplings to the supergravity fields
 (corresponding
to  contact terms
in the correlators).

Indeed, in type IIB string theory the $\N=4$
SYM theory appears (as  the leading term in the  Born-Infeld action) 
in the description 
of  D3-branes and thus, in  general, 
is coupled to  the fields of $D=10$ 
type IIB supergravity multiplet. 
The fields of the $D=4$ CSG  which naturally couple 
to $D=4$ SYM  theory 
may be  interpreted  as 
a particular truncation of the  $D=10$ type IIB multiplet.

We shall mostly concentrate on the term in the  SYM effective action $W$
which is quadratic in the conformal supergravity fields. 
The imaginary part of this    term 
 is related \gk\
 to the  classical absorption 
cross-section 
of certain type IIB  supergravity  modes  by D3-brane.
The conformal supergravity 
    gives a universal supersymmetric 
 description of different    marginal
(or `minimal', i.e.   dilaton, longitudinal graviton, etc.)
 cases discussed
  in \refs{\kl,\gkt}.
  Non-marginal  cases \refs{\hashi,\raj} (see also \deal) 
   are not directly 
described by  coupling of CSG to SYM. It is  natural to expect that  
 the conformal supergravity multiplet
 (supplemented by the anomalous  degrees of freedom)
can be coupled to the full D3-brane Born-Infeld action since the latter 
can be coupled to the type IIB multiplet.\foot{The BI  action 
for a D3-brane probe in a background of large $N$ D3-brane source
(or  in AdS$_5 \times S^5$ space) 
has (spontaneously broken) conformal invariance \refs{\mald,\kallo}
so it is  likely  that it  may be coupled  to the CSG fields
(viewed, e.g.,  as boundary values of the $D=5$ GSG fields).
Alternatively, this   $D=4$ BI action ($F^2 + { 1 \ov X^4} F^4 + ...$) 
 can be  interpreted  \refs{\chep,\malda}
as the leading large $N$ part of the quantum  SYM 
effective action  in the background with a non-zero value of the scalar
field $X$. If one includes 
  in addition the conformal supergravity background
 the resulting quantum  large $N$ SYM effective action 
 may  represent  the coupling of CSG to BI action.}

We shall show 
 that the leading 
gravitational term in $W$ (quadratic in the linearised Weyl tensor)
  does indeed 
emerge from the $D=5$  Einstein  action   when it is evaluated 
on  the solution of the Dirichlet problem in the AdS$_5$
space.
 Namely, we shall explicitly compute the  term bilinear 
 in the graviton perturbation in the  
 Einstein part of the  GSG  action,  
clarifying   some subtleties  (absent in simpler  cases of 
  scalar \refs{\gkp,\wit}
 and vector \wit\  perturbations)  involved in  
 establishing its  relation with the correlator of 
 the two energy-momentum
 tensors of SYM theory.\foot{One
 component 
 of this correlator was already  discussed in \gkp.
 Our aim will be to 
see  how the full  expression  
comes out 
of  the GSG action in a direct way.
}  In particular, we shall find that 
in order to guarantee the conformal invariance 
of the supergravity expression
one should 
start with  the $D=5$  action  containing an additional 
boundary (counter)term proportional to the volume of the boundary. 
The leading-order  solution of the Dirichlet problem for 
the graviton we shall find    allows  one 
to compute  also  the   3-point functions 
involving gravitons. In particular,  we shall 
determine the graviton-dilaton-dilaton  function 
 complementing the  recently obtained 3-scalar
\refs{\arv,\muck,\freed}
and 3-vector and scalar-scalar-vector  \freed\
 results.

In section 2  we shall  review  some aspects of
the SYM effective action in the  conformal supergravity background. 
In section 3  we shall discuss   the correspondence between the 
SYM effective action in the CSG background and the on-shell value 
of the $D=5$ gauged supergravity action suggested in \refs{\gkp,\wit} 
and its
possible tests.
In section 4 we shall compute the term bilinear in 
the gravitational
perturbation  in the $D=5$ Einstein action and
 demonstrate its  equivalence with the corresponding term
 in the  $D=4$  effective action.
In section 5  we shall apply the results of  
section 4 to compute the  dilaton-dilaton-graviton
term in the $D=5$ action   which is found to be consistent
with the  conformal invariance of the boundary theory.

\newsec{Effective action  of $\N=4, D=4$  SYM theory in   
$\N=4, D=4$ conformal
supergravity background}

The conformally invariant $\N=4$  super Yang-Mills  theory 
can be coupled in a natural way to $\N=4$  conformal supergravity  
(or Weyl)  multiplet. The action  describing  $\N=4$ SYM in  a background 
of $\N=4$ CSG  was  found at 
the linearised  level in \bdd\ and at  the full non-linear 
level  in \dero.
The   leading terms in the Lagrangian for a single $\N=4$ 
vector multiplet $(A_m, \psi_i, X_{ij})$
coupled in a $SU(1,1)$ covariant way
to the fields of  CSG 
(bosons $e^a_{m}, V^i_{mj}, \vp, T_{mn}^{ij}, E_{ij},
\D^{ij}_{kl}$    and fermions
 $\psi^i_m, \Lambda_i, \chi^{k}_{ij}$)\foot{In sections 2 and 3 
  $m,n=0,1,2,3$ are the  space-time
 and $i,j=1,2,3,4$ are  the $SU(4)$ indices.} 
can be written in the following schematic  form
$$
  L_{\rm SYM} = - \four (e^{-\p} F^{mn} F_{mn}   +   \C   F^{mn} F^*_{mn})  
-\ha \bar \psi^i \g^m D_m \psi_i 
- \four X_{ij}( -  D^2  + {\textstyle{1 \ov 6} } R) X^{ij}
$$\eqn\coup{
- \   X_{ij} F^{+mn} T_{mn}^{ij}
+    \D^{kl}_{ij}  (X^{ij} X_{kl} - \six   \d^i_k \d^j_l |X|^2) + ...
 + h.c. \ .  
}
Here the complex scalar  $\vp$ was set equal to $ \C + i e^{-\p}$
(this scalar is  present  also in  the $D=4, \N=4$ Poincare 
and $D=10,\N=2$ type IIB supergravity, and,   from the 
string theory point of
view,  is 
a combination of the dilaton and the  RR scalar).\foot{$SU(1,1)$ or $SL(2,R)$ 
 is a symmetry of the SYM-CSG 
coupling if the  transformation of the  CSG 
scalar $\vp$  is  accompaneed
by the duality rotation of the  SYM vectors \refs{\bdd,\dero}.
 This symmetry  should be   a manifest
off-shell symmetry of  the CSG action under which only the   
scalar $\vp$  is transforming. Let us recall  also that the 
   CSG theory in 10 dimensions \bed\ contains the  same 
fields  as  the (dual version of) $\N=1,D=10$ 
supergravity ($e^A_M, A_{M_1...M_6}, \Phi;  \psi_M, \chi$)
 but the real scalar $\Phi$  and  the 
Majorana spinor $\chi$  are subject to differential constraints.
 The  $D=4$ CSG scalar $\vp$  originates  upon dimensional reduction 
  from the 
  real $D=10$ scalar $\Phi$   and a  component of $A_{M_1...M_6}$.}
  
The resulting  action 
$I_\SYM(\A,\ge)=\int d^4 x \sqrt g L_{\SYM}$,   where $\A$ stands 
for the fields of the SYM multiplet and $\ge$ for the off-shell 
fields
of  the  CSG  multiplet,\foot{We assume that  $\ge$  
 include   also 
 the  conformal `gauge' or `anomalous' degrees 
of freedom (the conformal factor of the metric and its superpartners) 
which decouple at the classical level.}
 may  be viewed  as the $\N=4$ 
 locally superconformal  invariant  generalisation 
of the standard    coupling of the YM theory to the metric 
$\sqrt g g^{mk} g^{nl} 
F_{mn} F_{kl} $. 
 
Let us consider  the  SYM partition function 
in external CSG  background 
\eqn\gen{
Z(\ge) = e^{- W(\ge)} = \int d\A \ e^{- I_{\rm SYM}(\A,\ge)} \ . }
 The classical 
action  $ I_\SYM (\A,\ge) $   is invariant under  the 
local superconformal transformations
 but 
this is no longer true for the   effective action 
 $W(\ge)$ 
 (computed 
in a generally covariant and supersymmetric  way):
it will also depend on  the `anomalous' parts of the 
 CSG fields $\ge$.
 While  the 
$\N=4$ SYM theory  is finite in flat space, it  may still 
have  divergences  when coupled 
to external fields.\foot{Equivalently, the  correlators of composite 
operators may contain divergences.  
They  may  be defined  using  special (normal-ordering, etc.)
  prescriptions to be 
consistent with   conformal invariance
of the flat-space SYM theory,  but this  may  break  
 the general covariance of the SYM-CSG coupling.
The analogy with the  2d case is  only a  partial  one 
 since in $D =4$  the conformal group is finite-dimensional 
 and thus is very different from general covariance.}

 The divergent part of    $W$ 
which 
 is   superconformally invariant 
is nothing  but the  action of the  $\N=4$ conformal supergravity\foot{Here $\L$ is an
UV cutoff.
Because of
supersymmetry 
  the only
 non-vanishing divergence is
   logarithmic one.
 It  is natural to  include  the 
 total derivative (Euler density)  term $R^*R^*$  in the CSG action 
 since  then  the full divergence and anomaly 
 of the SYM theory is    determined simply   by the CSG action.
 The
  coefficient of the $D^2 R$ term in the conformal anomaly (see below)  is
 ambiguous. If one defines the Weyl tensor in 4 dimensions 
 (as we are assuming here) 
 this coefficient automatically cancels out for the $\N=4$ SYM multiplet.
 It is non-zero if one defines 
 the Weyl tensor in $4-\epsilon$ dimensions as in \duff.
 In that  case  one is to add the term
 ${2\ov 3} D^2 R$ to the $C^2 -R^*R^* = 2 (R^2_{mn} -
 {1\ov 3} R^2) $ combination in the CSG action.
 }
\eqn\www{
W= W_{\infty} + W_{\rm fin} \ , }
\eqn\div{
W_{\infty} = - \ \b\  \ln \L \   I_{\rm CSG} \ ,  \ \ \ \  \ \ \ 
  \b = { \n \ov 4 (4\pi)^2} \ , \ \ \   \ \ \   \n= N^2\ , 
}
\eqn\csg{I_{\rm CSG}
= \int d^4 x \sqrt g \ L_{\rm CSG}  
\ , } $$
\ \ \ 
L_{\rm CSG} =   C_{mnkl} C^{mnkl}  -  R^* R^*   + 
2 F^{imn}_{j } (V) F^j_{imn } (V)
$$
$$  + \  
 4[  D^2\vp^* D^2 \vp  - 2    (R^{mn} - \third g^{mn } R  ) D_m \vp^*
 D_n  \vp ]
$$ $$   + \
 16  (D^m T_{mp}^{ij+} D_n T_{ij}^{np-}  - \ha R^m_n T_{mp}^{ij+} T_{ij}^{np-}
)
 -   E^{ij}( - D^2  + \six R ) E_{ij}
 $$ $$+ ... +   \D^{ij}_{kl} \D^{kl}_{ij}
 + {\rm  fermionic\  terms}  \ . 
$$
The  kinetic terms for the fermions 
$\bar  \psi^i_m \Delta_{3 mn} \psi_{in}
+ \bar  \L^{i} \Delta_{3 } \L_{i}
 +  \bar  \chi^{k}_{ij} \g^m D_m \chi_{k}^{ij}, 
 $  
 like the kinetic terms for the bosons,
 contain the Weyl-invariant differential operators 
 $\Delta_p = (\g^m D_m)^p + ...\  $.\foot{The action of $\N=4$
 CSG was originally    found  only to the 
 quadratic order in the fields
\bdd.  Its   dependence  
on  the metric can be determined exactly 
from the condition of local Weyl invariance \refs{\ftt,\ft}.
The $SU(1,1)$ invariant version of $\N=4$ CSG  is the
`minimal' one, i.e. its action 
  cannot  contain 
scalar-vector  $h(\vp) F^2$ and  scalar-Weyl $ f(\vp) C^2$ 
couplings (which are not ruled out on the basis 
of Weyl invariance only). 
 Here we ignore higher-order terms in the fields (since the 
 explicit $SU(1,1)$ 
 invariant
 form
 of the scalar term is not known) and so that $\vp$  is assumed to be 
 equal  simply to $ \C - i\phi$.}
 \ \  $\n$ is the number of vector multiplets, i.e. 
 $N^2$
  in the $U(N)$ or large $N$ \   $SU(N)$   SYM case.
The UV divergence  \div\ is directly related to  the conformal
anomaly (assuming, of course,   that the  UV cutoff,  
e.g., 
$g_{mn} \Delta x^m \Delta x^n > \L^{-2}$, preserves general covariance),
\eqn\confa{
< T^m_m >  = {2 g_{mn}\ov \sqrt g}     {\delta W \ov \delta g_{mn}}
 = -  \b \   L_{\rm CSG}   \ . }
The one-loop  coefficient  here  can be  found 
by summing the  contributions  \duff\ of the  fields  of the SYM multiplet.
Since 
 the conformal anomaly is in the same  supersymmetry 
multiplet   with the axial  $SU(4)$ anomaly
\eqn\axi{
D_m {\delta W \ov \delta V_m} = \b \   F^{mn}(V) F^*_{mn}(V) \ , 
}
which should receive only the  one-loop contribution, it is natural 
to expect that in the present  $\N=4$ SYM case the above one-loop 
expression for $W_{\infty}$ \csg\  (and thus  for the anomalous 
part of $W$ discussed below) 
is  actually 
exact  to all loop orders  \refs{\fftt,\gk}.

The dependence of the finite part $W_{\rm fin}$
of the effective action  
$$ 
W_{\rm fin}  = W_{\rm anom} + W_{\rm inv} \ 
$$
 on the anomalous degrees of freedom 
can be determined  by integrating the anomaly relations
as in the 2d case \polya.
The  dependence of  $W_{\rm anom}$ on the conformal 
factor of the metric $\s$ 
 was found 
 in 
\refs{\fftt}.
This  part of the action (which is non-local when expressed
in terms of the original unconstrained metric)
takes the following explicit form\foot{We  set $g_{mn}= e^{2\sigma}
\td g_{mn},$ 
with $\td g$   subject to the  conformal gauge condition   \fv\ 
$R(\td g)=0$, i.e.  $e^\s = 1 - \six ( -D^2 + \six R)\inv R$.
We specify the expression in \fftt\ to the present $\N=4$ SYM
case where we do not introduce the $D^2 R$ term in the conformal
anomaly.  Equivalent
(but  corresponding  to a different 
  splitting 
  of the full $W$ into the anomalous and conformally-invariant parts)
    non-local expression 
for  $W_{\rm anom}$ 
  has  the form \rig\  \ 
$\int (C^2 - R^*R^* + { 2 \ov 3 } D^2 R +  F^2 + ... ) \Delta_4^{-1}
(R^*R^* - { 2 \ov 3 } D^2 R)  , 
$
  where $\Delta_4$ is the  4-th order conformally-invariant  
  scalar operator  \ftt, i.e. 
  the kinetic operator of $\vp$ in \csg. 
  }  
\eqn\ses{
W_{\rm anom} (g) =
-  2 \b \int d^4 x \sqrt { \td g} 
\bigg[ \big(\td R^2_{mn} -\third \td R^2  
+ 
  F^2_{mn} +  2 \td D^2 \vp^* \td D^2 \vp + ... )\ \s 
}
$$ +\  2 \td R^{mn} \del_m \s \del_n\s 
+ 2 \td D^m\s \td D_m\s \td D^2\s + (\td D^m\s \td D_m\s)^2 
 \bigg]
\ .   $$
Here  dots stand for terms depending on  other CSG 
 fields  (which  are  the same as in the CSG Lagrangian as can be  seen   
by replacing $\L$ by $ \L e^{\s}$  in the divergent part of the
effective action in \www).

Similarly, integrating the $SU(4)$ axial anomaly relation one finds  that
  $W_{\rm anom}$  
contains the term 
\eqn\axx{
W_{\rm anom} (V) = \b  \int 
     F^{kn}(V)  F^*_{kn}(V) \ D^{-2} D^m V_m \   + ...  \ . }
The  terms  quadratic in the field
strengths  (i.e. the leading terms  in the weak-field 
expansion)
 which are directly related to the divergent and conformal anomaly
parts of $W$  can be written  in the following covariant form\foot{Here
 $-D^2$ stands for  a 
Laplacian which  in general contains also curvature terms.
 Contributions  of higher than second order
  in curvatures are much more complicated
\vil. Similar quadratic   action
  was  also  discussed 
   in connection 
  with two-point and three-point correlation functions
  of  4d conformally invariant theories in \osb. }
$$
W_2 = \ha \b  \int   \bigg[ 
 C^{mnkl} \ln ( {-D^2 \ov \L^2}) C_{mnkl}  $$ 
 \eqn\qua{  + \  2 F^{mn}  \ln ( {-D^2 \ov \L^2}) F_{mn} 
+  4 D^2 \vp^* \ln ( {-D^2 \ov \L^2})  D^2 \vp + ... 
\bigg]  \  . 
}
For  $g_{mn}= e^{2\sigma}
\td g_{mn}$  the term in \qua\ which is  linear in $\s$ \ 
is indeed  the same as in \ses. 
As already mentioned above, 
the imaginary part of the quadratic ($p^{2n} \ln p^2$, \ $n=4,2$) 
term in this  effective action 
(or discontinuity of the 2-point  correlation function of the corresponding
operators in the SYM theory) 
is related to the classical D3-brane 
absorption \gk\ 
of dilatons,  
gravitons  and other `minimally' coupled CSG fields 
(or certain parts of the original  type IIB supergravity fields).

\newsec{$D=4$ Super Yang-Mills  -- $D=5$  gauged 
supergravity  relation}

According to the suggestion of \refs{\gkp,\wit},  in the large $N$,
 $g_{\YM}^2 N \gg 1$ limit  \mald\ of SYM theory 
one should have the following   equality  
\eqn\yyy{
W(\ge) =  I_{\rm GSG} [G (\ge)] \ 
}
between 
the  SYM effective action $W(\ge)$ in a  conformal 
supergravity background\foot{We  consider only  the 
coupling to  the marginal operators (and ignore also possible quantum
corrections on the supergravity side).
It is expected  \refs{\gkp,\hog,\wit} (see also \nensfet\ and refs. there)
that a similar relation  is true  also between correlators
of chiral  SYM operators and   terms in the 
 action of $D=10$ IIB supergravity
compactified on $S^5$  which correspond to 
the KK modes.}  and the action
 $I_{\rm GSG} [G (\ge)]$  of the $D=5,$ $\N=8$ 
gauged supergravity \gaug\ 
evaluated  on the classical solution with the boundary 
values of the $D=5$ fields $G$  being  related  to  the  $D=4$  CSG 
fields  $\ge$.\foot{For example, in 
addition to 15 vectors of $SU(4)$  the $\N=8,D=5$ 
GSG action 
contains  also 
6 self-dual antisymmetric tensors (dual to 12 vectors in ungauged case 
and having 
 kinetic terms which are of first order
in derivatives \gaug, 
 $ \ep^{\m\n\l\k\r}  B_{\m\n}^{ia} D_\k
B_{\l\r}^{ia}$)
    which are counterparts of the   6
self-dual antisymmetric tensors  $T^{ij}_{mn}$ of $D=4$ CSG theory.
}
As was pointed out  in \wit,  the IR divergences in the GSG action on 
\ads background 
are related to divergences and anomalies in the SYM effective. 
The UV cutoff  $\L$  in  SYM theory  is    related  to the 
distance to the boundary of \ads  boundary in  the picture of \wit\
(or   to  the radius of \ads in the picture of   \gkp\
which may be more natural  in the context of 
making  connection to the full D3-brane geometry).

The relation \yyy\  was   
 demonstrated   at the level of the  quadratic terms (2-point functions)
 in 
scalars \refs{\gkp,\wit} and 
  vectors \wit. 
The   scalar   $\int d^5 x  \sqrt g (\del_\m \p)^2 $
and vector $\int d^5 x \sqrt g  F^2_{\m\n}$ terms in the GSG action 
were  shown to lead to the boundary terms 
$\int d^4 x \  \p \del^4 \ln {-\del^2 \ov \L^2} \p$
and $\int   d^4 x \ F^{mn} \ln {-\del^2 \ov \L^2}  F_{mn}
   $ $=  - 2\int  d^4 x \  V^\perp_m  \del^2  \ln {-\del^2 \ov \L^2} 
     V^\perp_m,       $
in agreement with the  structure of   $W$\qua.

Below we shall consider   a similar test 
for the  term which is quadratic in the perturbation   of the metric.
It turns out that to ensure that only the transverse 
traceless part  of the  graviton 
 $\bar h^\perp _{mn}$ is coupled at the boundary
 one needs  to make a special choice 
of the boundary term in the GSG action.
  We shall  demonstrate  in section 4 that  the $D=5$  Einstein+cosmological+boundary
  term  expanded near AdS$_5$ background and evaluated on the solution
  of the Dirichlet problem to the $O(h^2)$ order
  reproduces  the  quadratic term in the  Weyl tensor  part of  \qua, i.e. 
  \eqn\weyl{
  C^{mnkl} \ln {- D^2 \ov \L^2}  C_{mnkl}  
   =    \ha \del^2 \bar h^\perp_{mn} \ln {-\del^2 \ov \L^2} \del^2 \bar
h^\perp_{mn} 
  + O(h^3) \   .} 
It remains an interesting problem to see how 
 the full non-linear expression  
   for the  local (divergent) $D=4$ CSG 
part \div\ of the SYM  effective action $W$   can 
emerge  from the  $D=5$ GSG  action.
 
The finite 
anomalous   part
$ W_{\rm anom}$  of the SYM action which is closely related to the local 
divergent part   $ W_{\infty}$ 
starts with terms  which are {\it cubic} in the fields \ses,\axx. 
These  should thus originate from certain 
cubic   terms  on the GSG side.
In particular, one  should be able to reproduce the  scalar 
term 
 $\int d^4 x \ \s \ \del^2 \vp^* \del^2 \vp $  in \ses\   by starting with 
the 
$\int d^5 x \sqrt g  g^{\m\n}  \del_\m \vp^* \del_\n \vp$ 
term in the GSG action  and 
replacing  the fields  by their  classical expressions on the
\ads 
background  with  the boundary conditions relating them to 
the $D=4$ 
fields. The same should be 
true for the vector terms
$\int d^5 x \sqrt g\  F^2_{\m\n}(A)  \to \int d^4 x \sqrt {\td g}
\ \s\  F^2_{mn}(V)$.

As was   pointed out in \wit,   the presence of  the  
Chern-Simons term 
$\int  A\wedge   dA \wedge dA $
in the $D=5$  GSG action \gaug\  is 
 related to  the $SU(4)$  anomaly \axi\  in the SYM action 
 (see also \freed).\foot{
 Indeed, the transformation
 $\delta A = d a$ 
gives the boundary term $ \int d^4 x\  \bar a F F^*$, 
where  $\bar a $ (having 
the meaning
of the  $SU(4)$ gauge parameter)
is the restriction of  $a$ to the boundary 
 and $V= A|_{\del M}$.}
  This  implies    that 
this   CS term  in the GSG action  evaluated 
on the  solution of the Dirichlet problem should reproduce 
the 3-point anomalous term \axx\ in $W$.

 Following the same logic in the 
 case of the conformal anomaly  
and considering  the  variation of the  $D=5$ GSG 
action under the Weyl transformation of the metric 
concentrated at the boundary we immediately reproduce
the $F^2_{mn} $  term in the 
$D=4$ conformal anomaly relation \confa\ 
(note that the Maxwell action is not
 conformally invariant in $D=5$).
Similar argument should apply  (though in  a less 
straightforward   way) also  to the scalar 
and graviton  terms\foot{The term linear in the Weyl  variation of the
$D=5$  Einstein term is  determined from 
$ (\sqrt g  R )' =  \sqrt g e^{(D-2)\s} [ R + 2 (D-1) \nabla^2 \s 
- (D-1) (D-2)  (\nabla \s)^2 ]$, 
where 
 $g'_{mn} = e^{2 \s} g_{mn}$.} in the $D=5$ GSG action:
expressing the $D=5$ fields in terms of their boundary values
one should get (to the second order in the scalar and graviton fields)
$\int d^5 x\ \s (\del_\m \p)^2 \to  \int d^4 x\ \s (\del^2 \p)^2$, 
\ 
$ \int d^5 x\  (\s \ R  +  {8 \ov 3} D^2 \s) 
 \to  \int d^4 x\ \s (\del^2 h^\perp_{mn}  )^2$.
We shall not discuss the  anomalous 3-point terms any  further 
in this paper.\foot{  To reproduce them from the $D=5$ action in a 
systematic way
  seems  to require to take into account some extra 
  boundary  contributions in the evaluation of the 3-point terms in the
  action which  result  from  `subleading' contributions
  to the leading-order  solution of the 
  Dirichlet problem for the  graviton discussed in the next section.}
 
The simplest  non-anomalous 3-point function involving graviton
is  the scalar-scalar-graviton one.\foot{Note that there are no
massless 
scalar-vector-vector couplings  both  in the $D=4$
CSG and  and the  $D=5$ GSG actions.   3-point terms involving  
vectors only and  vectors and massive KK scalars were systematically discussed in \freed. }
The term $ \int d^5 x \  \del_\m \p \del_\n \p h_{\m\n}$  in the GSG action 
should  reproduce, in particular,  the  divergent 
$\p \p h$ term in  $W$ contained in the 
$\vp \Delta_4 \vp$    part of the  CSG action
\csg.  We shall  compute this dilaton-dilaton-graviton function 
 in section 5.

\newsec{Metric  perturbations on AdS$_{d+1}$ background
and graviton  2-point  function}

\subsec{Notation and review of the scalar  2-point function}
We shall follow \wit\ and consider   the 
Anti de Sitter space of dimension $D=d+1$ with   Euclidean signature
and the (half-space) metric  
\eqn\adsm{
ds^{2} = g_{0\m\n} dx^\m dx^\n= {1 \over x^2_0} (d x^2_0  + d x_{i}^{2}) \ , \
\ \ \ \ \ \   
i=1,2,..., d \ .
}
In this 
hyperbolic space all conformal symmetries (e.g., scalings and inversions) are
isometries.
This  space has two boundaries:  $x_0 = \infty$, which is a single
point,  and  $x_0 =0$ which is $R^{d}$. The whole boundary is
topologically $S^{d}$. At $x_{0}=0$ boundary we will 
use the flat $R^{d}$ 
metric $\td g_{Bab}=\delta_{ab}$ 
which is conformally related to  $g_{0ij}$ \adsm.\foot{ 
This  flat metric is different from  the induced metric $g_{Bij}$ on the  
hypersurface $x_0=\ep\to 0$ close to the boundary, where  the
 total-derivative
part of the bulk  integrals appearing below will be   computed.}
The AdS$_{d+1}$  bulk  indices  will be 
denoted by $\m,\n, \a, \b, ...$ and will take values $ 0, 1,...,  d$.
We shall  use the notation  $x= (x_0, \vex)$, 
$\vex = (x_i)$, \ 
$i = 1, ..., d$.
The boundary indices  will be  labelled  by $a,b,... = 1, ...,  d$
(which, in general, 
  should be distinguished from   $i,j,...$). 
The bulk indices  will be  raised and lowered
by metric \adsm. Since the boundary metric is flat, 
   we  will  not distinguish  between  the 
upper and lower  boundary indices.

Let us start with a brief review (following  \wit)
of the calculation of the 
term bilinear in the scalar  field.
Considering  the  Euclidean  action for a massive scalar 
$
I (\p) = \ha \int d^{d+1} x \sqrt{g_0}\, [ (\del_\m \p)^{2} + m^{2} \p^{2}]
$
we are to solve 
$
D^{2} \p - m^{2} \p =0
$
with the   Dirichlet boundary 
condition $\p (x_0=0, \vex) =\p_0(\vex)$.  For  this
 we need to find  a  `propagator' $\KK$  which approaches 
a $\delta$-function at the boundary. 
One 
may choose  the  $\delta$-function
source to be  located at  $x_0 = \infty$.
Since both the boundary condition and the  metric  do not depend on
$\vex$, we can take $\p$ to be a function of $x_0$ only. Then  
the equation of motion  simplifies and the two independent 
solutions are:
$
\p(x_0)  = x_0^{\Delta_{\pm}},  \ \ 
\Delta_{\pm} \equiv  \ha (d \pm \sqrt{d^{2} + 4 m^{2}}).
$
The one which satisfies the boundary condition is 
$
\KK(x_0) \equiv \p (x_0)  = x_0^{\Delta_+}.
$   
Next,  we perform  the coordinate inversion 
which maps the point $x_0=\infty$
to the point $x=(x_0=0, \vec{x} = \vec x')$,
\eqn\inver{
x^{\mu} \ra {\hx^{\mu} \over f}, \ \ \ \ \ \ \ \ \ \
f \equiv |x-x'|^2=  x_{0}^{2} + \hx^{2}\ ,\ \ \ \ \ \ 
 \hx^{i}  \equiv  x^{i} - x'^{i}\ .
}
It is easy to see that \inver\  is  an isometry of AdS
space and maps the boundary to itself (at the 
boundary it induces a conformal transformation).
 After the inversion\ 
$
\KK (x_0) \to \KK(x_0, \vex; \vex') = ({x_0 \over f})^{\Delta_+}  . 
$
By superposition,  we  find the   general solution
of the Dirichlet problem 
\eqn\scas{
\p (x_0, \vex) =  c_{dm}  \int d^{d} x'\  {x_0^{\Delta_+}
 \over (x_0^{2} + |\vex-\vex'|^{2})^{\Delta_+}}\ 
\p_{0}(\vex')\ , 
}
where $c_{dm}= {\Gamma (\Delta_+) \over \pi^{d/2} \Gamma (\Delta_+ - {1 \ov 2} d)}.
$\  The  field 
$\p_{0}$ has the boundary theory interpretation of a
  scalar 
 with conformal dimension $d-\Delta_+$.
Plugging  the solution \scas\  into the action  $I(\p) $ 
we find  ($x_0=\ep\to 0$) \refs{\wit,\freed}
\eqn\soll{
I = - \ha  c_{dm}(2 \Delta_+ -d)   \int  d^{d} x \, d^{d} x' \, {\p_{0}(\vex)\  \p_{0}(\vex')
\over 
 (x^2_0 + |\vex - \vex'|^2)^{\Delta_+}}
\ . }
This  determines  the  two-point function of a conformal 
operator  ${\OO}$ with dimension $\Delta_+$, which couples to $\p_{0}$
(via $
\int d^{d}x \, \p_{0} {\OO}
$)   in the boundary theory.

\subsec{Quadratic term in  the gravitational   action }
Below we shall consider  the second-order term in the 
gravitational action expanded near AdS background.  
 The leading-order  solution of the Dirichlet 
 problem for the gravitational perturbation we shall 
find allows also to  find   the 
cubic terms in the action involving gravitons
(which determine the  3-point  correlators of conformal field theory
involving  the  energy-momentum  operator).
 One particular such term (dilaton-dilaton-graviton one)
 will be discussed in section 5.
 Given that the AdS space is a solution of the same 
  action one is perturbing, 
 the gravitational perturbation  case is  more  subtle  
 than the scalar 
 and vector  cases  considered in \refs{\gkp,\wit,\muck,\freed}.   
 
Our starting point is the Einstein action  with a  cosmological constant $\l$
in $d+1$ dimensions
\eqn\acto{
I=I_M + I_{\del M} \ , \ \ \ 
\ \   I_M = \int_M d^{d+1} x\ \sqrt g (R-2 \lambda)   \ , 
}
\eqn\bou{
 I_\dM = I^{(1)}_\dM  + I^{(2)}_\dM \ , \ \ \ \ \ \ 
 \ \ \ 
I^{(1)}_\dM = 2\int_{\dM} d^{d} x\ \sqrt{{g}_B}\  K = \ 2\ 
\del_{n} \int_{\dM} d^d x\ \sqrt{g_B} \ , }
\eqn\new{I^{(2)}_\dM = a \int_{\dM} d^{d} x\ \sqrt{{g}_B}  \ .
}
Here $g_B$ is the metric at the boundary
and  $K$ is the trace of  the extrinsic curvature of  the
 boundary, i.e. 
$I^{(1)}_\dM$ is the standard boundary 
term \refs{\hawkgib}.\foot{Choosing  a coordinate system at the boundary in
which 
$n^\a = (1,0,...,0)$ and 
 $\sqrt{g}= \sqrt{g_B}$,  we have
$
K = D_\a n^\a = {1 \over \sqrt{g_B}} \del_{\a}(\sqrt{g_B} n^\a)
={1 \over \sqrt{g_B}} \del_{n}\sqrt{g_B}, 
$ so that 
$
\int_{\dM} d^d x \sqrt{g_B} K 
=\int_{\dM} d^d x \ \del_{n} \sqrt{g_B} 
= \del_{n} \int_{\dM} d^d x \sqrt{g_B}\ .     
$}
Having the volume (cosmological) 
term  added to  the Einstein action, 
it is natural to introduce also the {\it additional} boundary 
term $I^{(2)}_\dM$
proportional to the area of the boundary.
The coefficient $a$ of this term will be chosen  so that 
to ensure the conformal invariance 
of the action  computed on the solution of the Dirichlet problem.

We shall expand the 
 metric near  the $M$=AdS$_{d+1}$  background
$
{g}_{\m \n} \to  g_{0\m \n} + h_{\m \n}, \,
$
i.e.  
$
{R}_{\m \n} \to  R_{0\mu\nu} + R_{1 \mu\nu} + R_{2 \mu\nu} + ...\ . 
$
The background  metric $g_{0\m\n} = x^{-2}_0 \delta_{\m\n}$
       satisfies
$
R_{0\mu\nu} = {1 \over d+ 1 }g_{0\mu\nu} R_0, \,\, 
R_0 = R= {2(d+1) \over d-1} \lambda,
$
where, as in \wit,  we  have  set  $\lambda=-\ha d (d-1),$
i.e. $R= - d (d+1)$.
To the first order in $h_{\m \n}$, the Einstein equations become
\eqn\ein{
R_{1\mu\nu}  + \ d \
   h_{\m \n} =0\ . }
Computing the   value of the Einstein Lagrangian 
to the second order in the  perturbation near the AdS$_{d+1}$ solution
we get 
\eqn\some{
\sqrt{{g}}({R} - 2 \lambda) =   - 2 d \   \sqrt { g_0}   
+  {\cal L}_{2} + 
 \sqrt{{g_0}} D_{\a}t^{\a}  + O(h^3)\ , 
}
where $D_\a t^\a$  represents the total derivative terms  ($D_\m$ is the
covariant derivative with respect to 
the background AdS  metric $g_0$ and $h^\m_\n \equiv
g^{\m\rho}_0 h_{\rho\nu}, \ h \equiv h^\m_\m$)
\eqn\tt{
t^{\a} = h^{\n \a}  D_\n h  + D^\a h^{\m \b} h_{\b \m}
- D_\m  (h^{\m \b} h_{\b}^{\a}) - \ha  h D^\a h 
+ \ha h D^\b  h_{\b}^{\a } -   D^\a  h  + D^\b h^{\a}_{\b}
\ .  }
 $\cL_2$ is the  action for a free graviton  in the 
AdS backgound,
  $$
{\cal L}_{2} = \ha \sqrt{g_0} \, \big[ \ha D_\m h D^\m h  
-  D_\m h  D^\n h^{\m}_{\n} + D_\m h^{\a \b} D_\a h^{\m}_{\b}
-\ha D_\m  h_{\a \b} D^\m h^{\a \b} $$
\eqn\dere{ +\  d \  (\ha h^{2}
-h_{\m}^{\n} h^{\m}_{\n}) \big]  \ . }
Using the equations of motion for  $h_{\m \n}$, we find 
that $\cL_2$ 
reduces to a total-derivative term 
\eqn\vet{
{\cal L}_{2} = \sqrt{g_0} D_{\a} v^{\a}
\ , \ \ \ \  \ \ \ \ 
v^{\a} = \four [h  D^\a h 
- D_\b (h h^{\a \b}) - h^{\m \n} D^\a h_{\m \n}
+ 2 h^{\m \n} D_\n h^{\a}_{\m }]  \ . }
At the boundary we  shall  choose the gauge 
in which  $h_0^0 = h_0^i=0$  so that 
$$
\sqrt{{g}_B} = \sqrt{g_{0B}} ( 1 + \ha \he
 - \four  h_i^j h_j^i + \eight
\he^2 + ...)\ , \ \ \ \ \   \he \equiv g^{ij}_0 h_{ij} 
 \ , 
   \ \ \ \ \ \sqrt{g_{0B}} = x_0^{-d} \ ,  $$
and  $ {\del \over \del n} = - x_0 {\del \over \del x_0}$.
Then
\eqn\kterm{
I_\dM^{(1)} = 2 \int_{\dM} d^{d}  x \ x^{1-d}_0 \bigg[ {d \over x_0}
(1 +  \ha \he -  \four  h_i^j h_j^i + \eight  \he^2 )  
-  \ha  ( \del_0 \he
- h_{i}^{j} \del_{0} h^{i}_{j} + \ha \he \del_0 \he) \bigg] \ ,  
}
and 
\eqn\bouterm{
I_\dM^{(2)} = a \int_{\dM} d^{d}  x \ x^{-d}_0  
(1 + \ha \he - \four  h_i^j h_j^i + \eight \he^2 ) \ . 
} 
Here (and below in all boundary integral expressions) 
$x_0 \equiv \ep \to 0$.

Rewriting    \tt,\vet\ in the explicit form, 
substituting
the resulting expressions into the action  $I_M$ \acto\ 
 and  combining its non-constant part 
 with $I_\dM^{(1)}$ \kterm\ and $I_\dM^{(2)}$ \bouterm\
we  find\foot{Note that 
$\int d^{d+1} x\  ( {\cal L}_{2} + \sqrt g_0  D_\a t^\a)
= \int d^d x  \sqrt { g_B} n^\a (v_\a  + t_\a) =
\int d^d x\  x_0^{1-d}   (v_0 + t_0)$.
The derivative terms  in  $I_{\del M}$ 
(the terms in the second 
parenthesis  in  \kterm) 
cancel the corresponding terms in $t_{0}$ (i.e. the
remaining derivative terms in $I$  are the same as in $v_0$).
 This is the expected effect of  the boundary term
$I^{(1)}_{\del M}$ (it should 
 compensate those terms in the variation of the Einstein action
which give rise to the normal derivatives
of the metric at the boundary).
} 
\eqn\htoact{
\eqalign{
I  & = I_{M} + I_\dM^{(1)} + I_\dM^{(2)}  = \int_{\dM} d^{d} x \  x^{1-d}_0 \ 
  \big( \four  h_{i}^{j} \del_{0} h^{i}_{j}  
- \ha h_{i}^{j} \del_{j} h^{i}_{0} ) \cr
& + [2 (d-1)  + a ] \int_{\dM} d^{d} x \  x^{-d}_0 \ 
(1 + \ha \he - \four  h_i^j h_j^i + \eight \he^2 )\ , \cr
} 
}
where we have   omitted  the terms 
proportional to  $h^0_0$ and $h^i_0$
which vanish at the boundary.
Fixing  the constant in  the boundary area  term  \new\ 
to be 
 $a = - 2 (d-1)$ so that 
 the second term in \htoact\
vanishes (implying, in particular, the vanishing 
of the volume divergence in the boundary theory),
  we  get  the following  
simple  result   for the quadratic term  in the action 
\eqn\toact{
I  = \int_{\dM} d^{d} x \  x^{1-d}_0 \ 
  \big( \four h_{i}^{j} \del_{0} h^{i}_{j}  
- \ha h_{i}^{j} \del_{j} h^{i}_{0} \big) \ . 
}
As we  shall  see  below, 
 \toact\  leads to  the  expected  conformally
invariant expression for the  graviton 2-point function.

\subsec{Solution of the Dirichlet problem and graviton 2-point function}
We would like to solve the Einstein equations
\ein\ with  the 
Dirichlet  conditions
at the boundary. We shall  use 
$h^{\m}_\n = g^{\m\rho}_0 h_{\rho\n} = x^2_0 h_{\m\n}$
as the basic variables (they are equal to the vielbein components
 of the metric perturbation in the present case of  the 
 conformally flat background metric $g_0$).
We shall assume that $h_{00}=h_{0i}=0$ at the boundary,\foot{Starting 
with nonvanishing $h_{00}, h_{0i}$ at the boundary
we can make a gauge transformation to set them equal to 
zero without affecting 
the value of $h^i_j$. }
i.e. impose the following boundary conditions
\eqn\bound{
h_{j}^{i}(x_0=0, \vec{x}) = \hh_{ab}(\vec{x})\  ,\ \ \ \ \ \ 
 \ \ \ 
h_{00} (x_0=0) = h_{0i}(x_0=0) =0 \ , 
} 
where $ h_{j}^{i} \equiv g^{jk}_0 h_{ki}$
and $\hh_{ab}$  is the  fixed boundary value of the metric perturbation. 
One  
 would like to find 
the solution which approaches
a  $\delta$-function at the boundary. 
As in the scalar case discussed above \wit, one 
  may first 
find  a  solution which approaches 
  $\delta$-function at  $x_0 = \infty$, i.e.  satisfies 
the  boundary condition 
\eqn\boundm{
h^i_{j}(x_0 \ra \infty, \vec{x}) \ra \infty \ , \ \ \ \ \ \ \ 
h_{00}(x_0 \ra \infty) = h_{0i}(x_0 \ra \infty)
 = 0 \ ,  
}
(implying also $h_{ij}(x_0 \ra \infty, \vec{x}) \ra \infty$)
 and then  use  the inversion transformation.
Again,  it is sufficient to    take $h_{\m \n}$  to be  a function
of $x_{0}$ only.
The  traceless part $\bar h_{ij}$
of $h_{ij}$  then  satisfies 
\eqn\heh{
\del^2_0 \bar h_{ij} -{d-5 \over x_0} \del_0 \bar h_{ij} - {2(d-2) \over
x_{0}^{2}}
\ \bar h_{ij} =0 \ ,  \ \ \ \ \ \ \ \
   g^{ij}_0 \bar h_{ij} =0 \ . 
}
The solution which vanishes at $x_0 =0$ and blows up at $x_0=\infty$ is 
$
h_{ij} \sim x_{0}^{d-2}  . 
$
The   equations for   the trace 
$
\he =  g^{ij}_0 h_{ij}
$  of $h_{ij}$ and $h_{00}$ are equivalent to 
 the constraint 
\eqn\Hzz{
h_{00} = -{1 \over d} (2 \he + x_{0} \del_0 \he) \ . 
}
It is 
straightforward to check  that 
$h_{0i}$ does  not couple to $h_{ij}$ and $h_{00}$. In view of 
\boundm, we can  consistently 
set $h_{00}$ and $h_{0i}$ to be zero everywhere. 
Then the only non-vanishing   solution of 
\Hzz\  is   $\he \sim x_0^{-2}$. This 
does not satisfy the boundary condition \boundm, 
 so  that  we should set $\he=0$.

The  solution of \ein,\boundm\   is thus 
\eqn\sol{  
h_{ij}(x) =\  \kappa_d\    x_{0}^{d-2} P_{ijab} \hh_{ab}\ , 
\ \ \ \ \ \ \ \ h_{00} = h_{0i} =0 \ , 
}
where  $\hh_{ab}$ is  an arbitrary 
tensor (which will be  related to 
the   perturbation of the  metric at the boundary),  
$\kappa_d$ is a normalization constant to be determined later,
and $P_{ijab}$ is  a traceless projector (recall that $g_0^{ij} = x^{2}_0
\delta^{ij}$;   the indices $a,b$ are contracted with flat metric)
\eqn\tenA{
P_{ijab}  =   \ha  (\d_{ia} \d_{jb} + \d_{ja} \d_{ib})
-{\textstyle{1 \over d}} \d_{ij} \d_{ab} \ . 
}
Performing now the inversion \inver\ ($x=(x_0,\vec x), x'=(x_0,\vec x')$)
$$
h_{\m \n}\  \ra \  f^{-2} J_{\m \r}(x-x') J_{\n \lambda}(x-x') \ 
h_{\rho \lam}(x)
\ , \ \ \ \ \ \   f\equiv |x-x'|^2 = x^2_0 + |\vec x - \vec x'|^2 \ , 
$$
with  $J_{\m \n}(x) $  defined by\foot{The indices of products 
of  $J_{\m\n}$  and $P_{ijab}$ will always be contracted with 
flat metric.}
\eqn\jten{
J_{\m \n}(x) = \d_{ \m \n} - {2 x_\m x_\n \over |x|^2}  \ , 
}
we transform  \sol\  into 
$$
h_{\m \n} = \kappa_d {x_0^{d-2} \over f^{d}} 
J_{\m i}(x-x') J_{\n j}(x-x') 
\  P_{ijab}\  \hh_{ab}( \vec x') \ . \ \ \ \
$$
The general solution  is  found by  superposition,  
\eqn\gensol{
h_{\m}^\n (x_0, \vec x) = \kappa_d \int d^{d}   x' \ {x_0^{d} \over f^{d}}\ 
J_{\m i}(x-x') J_{\n j}(x-x')
\ P_{ijab}\  \hh_{ab}( \vec x') \ . 
}
Since 
$$
\lim_{x_0 \ra 0} {c_d \  x^{d}_0 \over(x^2_0+ |\vec x -\vec
x'|^2)^{d}} =  \d^{(d)}
(\vec{x}- \vec x')  \ , \ \ \ \ \  
c_d= {\Gamma (d) \over \pi^{{d \over 2}} \Gamma ({d \over 2})}\ , 
$$
\eqn\deel{
\lim_{x_0 \ra 0} {x^{d}_0 \over(x^2_0+ |\vec x -\vec
x'|^2)^{d}} J_{ik}(x-x') J_{jl}(x-x') P_{klab} =
{d-1 \over d+1} \  c^{-1}_d \  P_{ijab} \   \d^{(d)}
 (\vec{x}- \vec x')  \  ,  
}
we set  
\eqn\kapp{
\kappa_d =   {d+1 \over d-1} c_d   \ ,  }
 to ensure that $h_{i}^{j}$ reduces to $\hh_{ab}$ at the boundary.

Alternatively,  we can 
perform  the  gauge transformation,
$$
h_{\m \n} \ra h_{\m \n} - \na_{\m} \eta_{\n} -  \na_{\n} \eta_{\m} \ , 
\ \ \ \ 
\eta_{\m} = - {x_0^{d-2} \over 4(d+1)  f^{d-1}} 
\del_\m J_{ij}(x-x') P_{ijab}\ \hh_{ab} \ . 
$$
to get a  more  `transparent'  expression for $h_i^j$,
\eqn\ansa{
h_i^j (x_0, \vec x) 
 = c_d  \int d^{d}  x' \  {x_0^{d} \over 
(x^2_0 + |\vec x - \vec x'|^2)^{d}
}\  P_{ijab} \  \hh_{ab} ( \vec x')\ , 
}
\eqn\ansb{ 
h_0^i(x_0, \vec x)  = { c_d \ d \over d-1} \int d^{d}  x' \ {x_0^{d-1} \over 
(x^2_0 + |\vec x - \vec x'|^2)^{d}
}\ 
B_{iab}(x-x')\   \hh_{ab}(\vec x')  \ , }
\eqn\anms{  
h_0^0(x_0, \vec x)  =  - { c_d \ d \over d-1} \int d^{d}  x' \  {x_0^{d} \over 
(x^2_0 + |\vec x - \vec x'|^2)^{d}
} \ 
C_{ab}(x-x')\   \hh_{ab} (\vec x') \ , 
}
where 
$$
B_{iab} \equiv  \four  \del_{i} J_{jk} (x-x') \, P_{jkab}
\  , \ \ \ \ \ \  
C_{ab} \equiv  J_{ij}(x-x') P_{ijab} \ . 
$$
Since $h^{i}_{j}$  approaches  $\hh_{ab}$ for $x_0 \to 0$
(while 
$h^i_0,h^0_0 \ \to 0$), $\hh_{ab}$ should thus  
 couple  to the energy momentum tensor of
the  boundary conformal field theory.\foot{
That $h^{i}_{j}$  is the relevant  variable  
 follows also 
from  expanding the  Born-Infeld action of a D3-brane probe 
in the  AdS$_5$  background.} 
 The consistency  of this interpretation 
 can be  confirmed by making a scale transformation or inversion in  the AdS
 space 
 (which  are  isometries  
 in the bulk,  inducing   conformal transformations 
at the boundary).     
 It is easy to check  using   \gensol--\anms\ (and following  the method
 described in \freed) that the induced transformation 
on $\hh_{ab}$ indeed coincides  with 
the  conformal transformation of the  graviton field  in $R^d$.

Using   \ansa,\ansb\  we find  
\eqn\basca{
h_{i}^{j} \del_{0} h^{i}_{j} = \ c_d^2\    d\  x_0^{d-1}\ \int d^{d} 
 x' d ^{d} x'' \ 
{x_0^{d} \over |x -x'|^{2d}}\  {\hh_{ab}(\vec x') P_{abcd}  \hh_{cd}(\vec x'')
\over |x -x''|^{2d}}  \ , 
}
$$
h_{i}^{j} \del_{j} h^{i}_{0} =-
 {  c_d^2\ d \over d-1}\ x_0^{d-1} \
\int d^{d}  x' d^{d}  x'' \ 
{x_0^{d} \over |x -x'|^{2d}}\  {\hh_{ab}(\vec x') X_{abcd}(x-x') 
\hh_{cd}(\vec x'')
\over |x -x''|^{2d}} \ , 
$$
where $
|x -x'|^2 \equiv  x^2_0 + |\vec{x}- \vec x'|^2   
$  and  $X$ is defined by 
$$
X_{abcd}(x) = P_{abcd} - {d+1 \over 2|x|^2} (\d_{bc} x_a x_d + \d_{bd} x_a x_c
+ \d_{ac} x_b x_d + \d_{ad} x_b x_c) + {2(d+1) \over |x|^4} x_a x_b x_c x_d  \
. 
$$
Substituting these  expressions into    \toact, 
we finally obtain the following  result 
 for the quadratic part of the action 
\eqn\twop{
I = \ b_d \  \int d^{d}  x' d^{d}  x'' \ 
{\hh_{ab}(\vec x') H_{abcd}(x'-x'')  \hh_{cd}(\vec x'')
\over |x' -x''|^{2d}}  \ ,  
}
where   $ b_d \equiv {d \over 4} \kappa_d =  {d(d+1) \over 4(d-1)} \ c_d$\   and 
$$
H_{abcd}(x)  \equiv \ha (J_{ac}J_{bd} + J_{ad} J_{bc}) -{\textstyle {1 
\over d}}
\d_{ab} \d_{cd}  $$
\eqn\deff{= \  P_{abcd} - {1 \over |x|^2} (\d_{bc} x_a x_d +
 \d_{bd} x_a x_c
+ \d_{ac} x_b x_d + \d_{ad} x_b x_c) 
+ {4 \over |x|^4} x_a x_b x_c x_d  \ . 
}
The  kernel in \twop\ (with $x_0 \to 0$, \ 
$|x|^2 = x^2_0 + |\vec x|^2 \to |\vec x|^2$)
has precisely the    structure  
expected  for the  correlator  of the  two 
energy-momentum tensor operators 
 in a conformally invariant theory
 (see, e.g., \refs{\frap,\osb}).  

We have obtained  \twop\ by  starting with  the  full expression
for the $D=d+1$  gravitational  action \acto\ 
  which includes  the standard 
boundary 
term \bou\ and the  boundary  area 
counter-term \new. 
 Since 
the quadratic action \dere\ for a free graviton in AdS space 
is invariant under the conformal transformations,
 it is  natural 
to  expect, at the same time,   that one should get a conformally invariant 
expression by just starting with $\cL_2$ alone \dere,\vet. This is indeed
the case: plugging  \gensol\ into \vet, 
one  finds  \twop\  (though with a different  normalization factor). 
A  non-trivial  issue  (which is important for 
the duality  \mald\ between the  type IIB  supergravity 
 on $AdS_5 \times S^5$ 
and $\N=4$ SYM) 
 is  the agreement of the  normalization factors. 
We expect that 
the correct normalization  of  anomaly-related terms
is obtained only if one starts   with the action 
 that  includes all  relevant 
boundary terms (with the condition of cancellation 
of the boundary  power divergence  
used to fix their relative coefficients).
At the same time, the boundary terms should not be relevant 
for the conformally-invariant 
higher-point functions  given by the bulk integrals.

Specifying to the case of $D=5$ ($d=4$) 
and  using   the momentum representation  \refs{\gkp,\arv}
we  
conclude  that  \twop\   coincides 
with 
  the  quadratic term \weyl\  
in the    effective  action 
\qua\ 
(with the $D=5$ IR cutoff $x_0=\ep\to 0$ 
 being  related to the $d=4$  UV cutoff $\L$ in 
\div\ and $h_{mn} \sim \hh_{ab}$).

\newsec{ Graviton-dilaton-dilaton  3-point function } 
Our aim here will be to compute  the  $h\p\p$  part of the term 
($\p$ is a massless  scalar) 
\eqn\sea{
I = \ha \int d^{d+1} x\ \sqrt{g} g^{\m \n} \del_\m \p
\del_\n \p  \  
}
in the 
action  of the $D=d+1$ dimensional gauged 
supergravity theory  on the solution of the Dirichlet problem 
in AdS$_{d+1}$
background
and  demonstrate the agreement with the 
expected  form   of the corresponding 3-point function 
in the  boundary conformal theory. 

Let us first  consider a simpler example of interacting scalar theory 
(see also \refs{\arv,\muck}):
$
I(\p) = \int d^{d+1 } x \ \sqrt{g_0} \ [\ha \del^\m \p \del_\m \p +  V(\p)] 
\ .
  $
The equation of motion with the Dirichlet boundary condition 
$$
D^{2} \p =  V'(\p)\  , \,\, \,\,\,\, \ \ \  \  \ \ 
\p (x_0=0, \vec{x}) = \p_{0}(\vec{x}) \
$$
is solved, to  leading order in perturbation theory,  by 
$$\p = \p_{1} +   \p_{2}\ ,\ \ \ \ \ \ \
D^2  \p_{1} =0\  , \,\,    \ \ \ \ \ \ \
D^2  \p_{2} = V'(\p_{1})\  , 
$$
where
\eqn\corr{
\p_{1} = \int d^{d}  x'\   \KK (x_0, \vec{x}; \vec x' )\ \p_{0} (\vec x')
 \ , 
\ \ \ \ \ \ 
\p_{2} = \int  d x_0' d^d  x'\  G(x_0, \vec{x}; x_0', \vec x' ) \ 
V'(\p_{1} (\vec x')) \  . 
}
Here $\KK(x,x')= \KK(x_0, \vec{x}; \vec x' )$ is the  `propagator'
introduced  in  section 4.1   and $G(x,x')= G(x_0, \vec{x}; x_0', \vec x' )$
is the  standard  Dirichlet Green function, i.e. 
\eqn\green{
 \KK(x_0, \vec{x}; \vec x' ) = \ \sqrt{g_{B}}\  n^\a \del_\a 
G(x_0, \vec{x}; x_0 '  \ra 0,\,  \vec x' ) \ , \ \ \ \ \ \ \ 
G(x_0 \ra 0, \vec{x}; x_0', \vec x' ) \ra 0 \ . 
}
Substituting the solution \corr\ 
into the action $I(\p)$,  we find, to the leading order 
in $V$,  
$$
I(\p) = \int d^{d+1} x \  \sqrt{g_0}\  [ \ha \del^\m \p_{1} \del_\m \p_{1} + 
 \del^\m  \p_{1} \del_\m  \p_{2}  +    V(\p_{1}) ]= I_{0} +   I_{1}  \ , 
$$
where $
I_{0} = \ha \int d^{d+1} x \sqrt{g_0}  \del^\m  \p_{1} \del_\m \p_{1} 
$
was computed  in section 4.1 (eq.\soll\ with $m=0$) and 
\eqn\sone{
I_{1} 
=  \int_{\dM}  d^{d} x\ \sqrt{g_{0B}} \  \p_{2} \ n^\a \del_\a  \p_{1}\  +  
\int_{ M} d^{d+1 } x\ \sqrt{g_0} \  V(\p_{1}) 
= \  \int_{M} d^{d+1 } x\        \sqrt{g_0} \ V(\p_{1})  \ ,  } 
where we have used that  $\p_{2}|_{\dM} =0$.\foot{Since the 
 boundary condition is saturated  by $\p_{1}$, the correction 
$\p_{2}$ should go to zero at the boundary.
In  the case of a compact source,  this  condition is satisfied
automatically  as  a consequence of   \green.
Since the source $V'(\p_{1})$ 
does not vanish at the boundary,  \corr\  may have an  extra boundary
contribution.
Here we  shall  ignore this problem.}
Thus to  find  the lowest order scalar  3-point function of the 
 we  need only to plug the free-field  Dirichlet problem 
 solution $\p_1$  into  the  potential term in the action 
 \sone.


Let us now return to   our problem  \sea. 
Taking  
$g_{\m \n} = g_{0\m \n} +  h_{\m \n}$,  expanding   
to the   lowest order in $h_{\m \n}$ and following the same steps as in the 
pure scalar case, we  find 
\eqn\aca{
I_1 = - \ha  \int d^{d+1} x\  \sqrt{g_0} \ h_{ \m}^\n T_\n^\m \ , 
}
where $T_{\n}^{\m}$ is the  energy-momentum tensor 
of the dilaton field in  the AdS$_{d+1}$   background 
$$
T_\n^\m  =g_0^{\m \s} \del_\s \p \del_\n \p -
 \ha \d^\m_\n g_0^{\r \s} \del_\r \p  
\del_\s \p  \ . 
$$
The solutions for $\p$  and $h^\m_\n$  are given by \scas\ 
(with $m=0, \ \Delta_+=d$) and \gensol, i.e. 
$$ 
\p(x)  = c_d  \int d^d y \ \KK (x,y)\  \p_0 (\vec{y})\ ,   
\ \ \ \ \ \ \   \KK(x,y) =  {x^d_0 \over |x-y|^{2d}}\ , 
$$
\eqn\cop{
h_\m^\n (x)  = \k_d \int d^d y  \ \KK(x,y)\
J_{\m i}(x-y) J_{\n j} (x-y)\ P_{ijab} \, \hh_{ab}(\vec{y})\ , } 
with $x=(0, \vec{x})$,\ $y=(0, \vec{y})$, \  $c_d= {\Gamma (d) \over \pi^{{d \over 2}} \Gamma ({d \over 2})}$, 
$\k_d = { d+1 \ov d-1 } c_d$ and  $J_{\m \n}(x) $  defined  in \jten.
Then  \aca\ takes the form  
\eqn\yyt{
I_1 =   \int d^d {x} \ d^d  {y} \ d^d  {z} 
\    A_{ab} (\vec{x},\vec{y},\vec{z}) \ \hh_{ab} (\vec{x})\ 
\p_0 (\vec{y})\  \p_0 (\vec{z}) \  , } 
with $$A_{ab}  \equiv  - \ha  \k_d  c_d^2 (A_{1ij}  - \ha A_{2ij}) P_{ijab}  \ ,  $$
$$
\eqalign{
A_{1ij} (\vec{x},\vec{y},\vec{z})& = \int {d w_0 d^d {w} \over w_0^{d+1}} \
\KK(w,x) \ 
J_{\m i}(w-x) J_{\n j} (w-x)\  w_0^2 \ \del_{\m} \KK(w,y)\ 
\del_{\n} \KK(w,z) \ , \cr
A_{2ij}(\vec{x},\vec{y},\vec{z})  & = \int {d w_0 d^d  {w} \over w_0^{d+1}} \
\KK(w,x) \ 
J_{\m i}(w-x) J_{\m j} (w-x) \ w_0^2\   \del_{\r} \KK(w,y)\ 
\del_{\r} \KK(w,z)\ ,  \cr
}  
$$
where the  repeated indices are contracted with flat metric 
and all  derivatives are  with respect to $w$.
Using the technique developed in \freed, it is straightforward 
to evaluate the above integrals. 
We find  
that  $A_2 =0$ and
$$
A_{1ij}  = - {2 \pi^{d \over 2}\  d^2\  [ \Ga ({d \over 2} +1 )]^3
\over \Ga (d+1)  \Ga (d+2)}\  \ 
{1 \over |\vec{x} - \vec{y}|^{d-2}}
{1 \over |\vec{x} - \vec{z}|^{d-2}}
{1 \over |\vec{z} - \vec{y}|^{d+2}} $$
\eqn\aaa{ \times \bigg[ 
({ x_i - y_i \over |\vec{x} - \vec{y}|^2} -
{ x_i - z_i \over |\vec{x} - \vec{z}|^2})
({ x_j - y_j \over |\vec{x} - \vec{y}|^2} -
{ x_j - z_j \over |\vec{x} - \vec{z}|^2})
- {{ 1 \ov d}} \d_{ij} {  |\vec{z} - \vec{y}|^2 \over
|\vec{x} - \vec{y}|^2 |\vec{x} - \vec{z}|^2} \bigg] 
 \ . }
If we define 
$$
  \lambda_{ab}^{\vec x}(\vec y,\vec z)   =\lambda_a^{\vec x}(\vec y,\vec z)
   \lambda_b^{\vec x}(\vec y,\vec z)
- {{ 1 \ov d}} \d_{ab} \lambda_c^{\vec x}(\vec y,\vec z)
 \lambda_c^{\vec x}(\vec y,\vec z)\ , \  \ \ \ \ 
\lambda_a^{\vec x}(\vec y,\vec z) = { x_a - y_a \over |\vec{x} - \vec{y}|^2} -
{ x_a - z_a \over |\vec{x} - \vec{z}|^2}\ , 
$$
then  the kernel in \yyt\  takes the form 
\eqn\voot{
A_{ab}(\vec x, \vec y, \vec z)  = {d^3 \, \Ga (d-1)  \over 8 \pi^d }
\lambda_{ab}^{\vec x}(\vec y,\vec z) 
{1 \over |\vec{x} - \vec{y}|^{d-2}}
{1 \over |\vec{x} - \vec{z}|^{d-2}}
{1 \over |\vec{z} - \vec{y}|^{d+2}} 
\ . } 
$A_{ab}$ has  precisely the  same  structure 
as the  conformally-invariant   correlation function
of the  three composite  conformal operators  coupled to  
graviton and  two  massless scalars
 (see, e.g.,  \frap, p. 104). 
 
 
 The conformal Ward identity requires  \frap:
\eqn\ward{
<T_{ab} (\vec x) \OO(\vec y) \OO(\vec z)> = 
-  {d \Delta_+ \over d-1} 
{\Gamma ({d \over 2}) \over 2 \pi^{d/2}} 
\lambda_{ab}^{\vec x}(\vec y,\vec z) \  ({|\vec{z} - \vec{y}| \over 
|\vec{x} - \vec{y}| |\vec{x} - \vec{z}|})^{{d-2 }}
\
<\OO (\vec y) \OO(\vec z)> ,  
}
where $\Delta_+$ is the conformal dimension of a scalar operator  $\OO$ 
(see section 4).\foot{The standard   differential form of  the Ward identity 
is
$$
\del_{a} <T_{ab} \OO_1 ... \OO_m> = - \sum_{k=1}^m [ \d (x-x_k)
\del^{x_k}_{b} + ...] <\OO_1 ... \OO_m>
$$
where the dots in the square  brackets  denote  possible contact
terms.  
For   the 3-point function, the conformal invariance requires the correlator
to take the form  (5.10)  up to an overall constant. 
To find the constant, we have  first to regularize (5.10),
and then take the derivative  over  $x$. 
Then the Ward identity in differential form given above will 
determine the overall constant to be the one  in (5.10). 
The differential form of  (5.10) is
$$
\del^x_{a} <T_{ab}(x) \OO_1(y) \OO_2(z)> 
= - \bigg[ \d(x-y) \del^y_b+  \d(x-z) \del^z_b
- {\Delta_+\over d} [ \del^x_{b} \d (x-y) +   \del^x_b  \d (x-z)] \bigg]
<\OO_1 \OO_2>.
$$
Note that the contact terms are also determined from
the  conformal symmetry.} 
In  the  case of the operator coupled to the 
massless dilaton ($\Delta_+ =d$), 
assuming that the coupling at the boundary of AdS$_5$ 
is given by $
 \int d^d x ( \ha  \hh_{ab} T_{ab} +  \p_0 \OO),
$
we  get
\eqn\enma{
<T_{ab} (\vec x) \OO (\vec y) \OO(\vec z) > = 4 A_{ab} (\vec x, \vec y, \vec z)
\  . 
}
It is easy to check that  with $A_{ab}$ given in 
\voot, the relation  \ward\ is  indeed satisfied.

The above  result for the graviton-scalar-scalar function can be 
generalised to the massive scalar case and is again found to  be  consistent 
with the Ward identity \ward.
 For 
 a massive scalar   
$
T_\n^\m  =g_0^{\m \s} \del_\s \p \del_\n \p -
 \ha \d^\m_\n (g_0^{\r \s} \del_\r \p  \del_\s \p 
+ m^2 \p^2 )  , 
$
and $A_{ab}$  in \yyt\  is given by 
\eqn\hht{
A_{ab}  \equiv  - \ha  \k_d  c_{dm}^2 
[A_{1ij}  - \ha (A_{2ij} + A_{3ij})] P_{ijab}  \ ,  
}
with
$$
\eqalign{
A_{1ij} (\vec{x},\vec{y},\vec{z})& = \int {d w_0 d^d {w} \over w_0^{d+1}} \
\KK(w,x) \ 
J_{\m i}(w-x) J_{\n j} (w-x)\  w_0^2 \ \del_{\m} \KK_{\Delta_+}(w,y)\ 
\del_{\n} \KK_{\Delta_+}(w,z) \ , \cr
A_{2ij}(\vec{x},\vec{y},\vec{z})  & = \int {d w_0 d^d  {w} \over w_0^{d+1}} \
\KK(w,x) \ 
J_{\m i}(w-x) J_{\m j} (w-x) \ w_0^2\   \del_{\r} \KK_{\Delta_+}(w,y)\ 
\del_{\r} \KK_{\Delta_+}(w,z)\ ,  \cr
A_{3ij}(\vec{x},\vec{y},\vec{z})  & = m^2 
\int {d w_0 d^d  {w} \over w_0^{d+1}} \
\KK(w,x) \ 
J_{\m i}(w-x) J_{\m j} (w-x)     \KK_{\Delta_+}(w,y)\ 
\KK_{\Delta_+}(w,z)\ ,  \cr
}  
$$
where we have used \scas\ and 
$
\KK_{\Delta_+}(w,z) = ( {
w_0 \over |w-z|^2})^{\Delta_+}.
$
 Since 
$
\Delta_+ (\Delta_+ - d) = m^2
$
 we find that $A_2 + A_3 =0$, 
while 
$$
A_{1ij} P_{ijab} = - \gamma  \lambda_{ab}^{\vec x}(\vec y,\vec z)\ 
({|\vec{z} - \vec{y}| \over 
|\vec{x} - \vec{y}| |\vec{x} - \vec{z}|})^{{d-2 }}\ 
{1 \over |\vec{z} - \vec{y}|^{2 \De_+}} \ , 
$$
\eqn\vvb{
\gamma \equiv  { 2\pi^{d \over 2} \De_+ ( \De_+ -\ha  d) \Ga ({ \De_+ -\ha d }) 
 [ \Ga ({d \over 2} +1 )]^2 
\over \Ga (\De_+) \Ga (d+2)} \ . }
Observing that 
$$
2 \gamma  \, \kappa_d c_{dm}^2
=  {d \Delta_+  \over d-1}  {\Gamma ({d \over 2}) \over  \pi^{{d \over 2}}}
( \De_+ - \ha d)\ c_{dm} \ , 
$$
 it follows from \soll\  that the Ward identity \ward\
is also satisfied in the massive case. 
Note that this is an independent confirmation that the normalization \freed\ 
 in \soll\  is the consistent  one.

\newsec{Concluding remarks}

We have thus established the  explicit 
 relation between the quadratic term in the 
$D=5$ Einstein action 
in the AdS$_5$  background and the quadratic term 
in the  Weyl tensor squared  part of the 
quantum effective action of the $D=4$ SYM theory 
in the conformal supergravity background.

As was already mentioned above, 
it would be very interesting to understand, in particular, 
how to extend the  equivalence between 
  the logarithmically  IR singular
part of the $D=5$ gauged supergravity action 
evaluated on  the solution of the Dirichlet 
problem  and the $D=4$ 
conformal supergravity action  \div,\csg\ 
 to the full non-linear level, and thus `derive'
 the $D=4$ conformal supergravity  (Weyl +...) action from the 
 $D=5$ gauged supergravity (Einstein +...) action.
 
A simple test of this   correspondence  between  the two  gravitational 
actions beyond quadratic level   would be   to check   explicitly 
that the  divergent  part of the 3-point function \yyt\
is indeed in agreement with the $h\p\p$ term in the  $\vp^* \Delta_4 \vp$
part of the CSG action
\csg.

\bigskip\bigskip

\centerline {\ \bf Acknowledgments}
\bigskip
We acknowledge the support 
 of PPARC,   the European
Commission TMR programme grant ERBFMRX-CT96-0045
and the NSF grant PHY94-07194.
A.A.T.  is  grateful  to   I. Chepelev,  M. Douglas,   M. Green 
and R. Metsaev 
for stimulating  discussions. H.L. would like to thank A. Matusis
for  useful correspondence.


\vfill\eject
\listrefs
\end